\documentclass[usenatbib,referee]{mn2e}
\usepackage{epsfig}
\usepackage{graphicx}
\usepackage{amsmath}

\def\beq{\begin{equation}}
\def\eeq{\end{equation}}
\def\bey{\begin{eqnarray}}
\def\eey{\end{eqnarray}}

\def\mpc{\, h^{-1}{\rm {Mpc}}}

\def\kpc{\, h^{-1}{\rm {kpc}}}

\def\kms{\,{\rm {km\, s^{-1}}}}

\def\msun{\, h^{-1}{\rm M_\odot}}

\def\gs{\mathrel{\raise1.16pt\hbox{$>$}\kern-7.0pt
\lower3.06pt\hbox{{$\scriptstyle \sim$}}}}
\def\ls{\mathrel{\raise1.16pt\hbox{$<$}\kern-7.0pt
\lower3.06pt\hbox{{$\scriptstyle \sim$}}}}
\def\gtsima{$\; \buildrel > \over \sim \;$}
\def\ltsima{$\; \buildrel < \over \sim \;$}
\def\prosima{$\; \buildrel \propto \over \sim \;$}
\def\gsim{\lower.5ex\hbox{\gtsima}}
\def\lsim{\lower.5ex\hbox{\ltsima}}
\def\simgt{\lower.5ex\hbox{\gtsima}}
\def\simlt{\lower.5ex\hbox{\ltsima}}
\def\simpr{\lower.5ex\hbox{\prosima}}
\def\la{\lsim}
\def\ga{\gsim}

\newcommand{\etal}{{et al.~}}

\title[Reconstructing the Cosmic Velocity and Tidal Fields]
        {Reconstructing the Cosmic Velocity and Tidal Fields with
          Galaxy Groups Selected from the Sloan Digital Sky Survey }
\author[Huiyuan Wang et al.]
   {\parbox[t]{\textwidth}{
       Huiyuan Wang$^{1,2}$\thanks{E-mail: whywang@mail.ustc.edu.cn},
       H.J. Mo$^{3}$,
       Xiaohu Yang$^{4}$
       and
       Frank C. van den Bosch$^{5}$
}\\
            $^1$Key Laboratory for Research in Galaxies and Cosmology, University of Science and Technology of China, Hefei, Anhui 230026, China\\
            $^2$Department of Astronomy, University of Science and Technology of China, Hefei, Anhui 230026, China\\
            $^3$Department of Astronomy, University of Massachusetts, Amherst MA 01003-9305, USA\\
            $^4$Key Laboratory for Research in Galaxies and Cosmology, Shanghai Astronomical Observatory, Shanghai 200030, China\\
            $^5$Astronomy Department, Yale University, P.O. Box 208101, New Haven, CT 06520-8101, USA
}
\date{
Accepted ........ Received .......; in original form ......}
\pubyear{2011}

\begin{document}
\maketitle \label{firstpage}
\begin{abstract}
  Cosmic velocity and tidal fields are important for the understanding
  of the cosmic web and the environments of galaxies, and can also be
  used to constrain cosmology.  In this paper, we reconstruct these
  two fields in the Sloan Digital Sky Survey (SDSS) volume from dark
  matter halos represented by galaxy groups.  Detailed mock catalogues
  are used to test the reliability of our method against uncertainties
  arising from redshift distortions, survey boundaries, and false
  identifications of groups by our group finder.  We find that both
  the velocity and tidal fields, smoothed on a scale of $\sim 2\mpc$,
  can be reliably reconstructed in the inner region ($\sim 66\%$) of
  the survey volume.  The reconstructed tidal field is used to split
  the cosmic web into four categories: clusters, filaments, sheets,
  and voids, depending on the sign of the eigenvalues of local tidal
  tensor.  The reconstructed velocity field nicely shows how the flows
  are diverging from the centers of voids, and converging onto
  clusters, while sheets and filaments have flows that are convergent
  along one and two directions, respectively.  We use the
  reconstructed velocity field and the Zel'dovich approximation to
  predict the mass density field in the SDSS volume as function of
  redshift, and find that the mass distribution closely follows the
  galaxy distribution even on small scales.  We find a large-scale
  bulk flow of about $117\kms$ in a very large volume, equivalent to a
  sphere with a radius of $\sim 170\mpc$, which seems to be produced
  by the massive structures associated with the SDSS Great
  Wall. Finally, we discuss potential applications of our
  reconstruction to study the environmental effects of galaxy
  formation, to generate initial conditions for simulations of the
  local Universe, and to constrain cosmological models. The velocity,
  tidal and density fields in the SDSS volume, specified on a
  Cartesian grid with a spatial resolution of $\sim 700\kpc$, are
  available from the authors upon request.
\end{abstract}


\section{Introduction}
\label{sec_intro}

The study of the large-scale structure in the universe traditionally
relies on large redshift surveys of galaxies, such as the Sloan
Digital Sky Survey (hereafter SDSS; York et al.  2000). However, since
galaxies are biased tracers of the mass distribution, one has to
understand the relationship between galaxies and dark matter before
using the galaxy distribution in space to study the mass distribution
in the universe. In recent years, tremendous amounts of effort have
been put into the establishment of the relationship between galaxies
and dark matter halos (the virialized clumps of dark matter
clumps). Since the relationship between the distribution of dark
matter halos and the mass distribution is well understood (e.g. Mo \&
White 1996; Jing 1998; Sheth \& Tormen 1999; Sheth, Mo \& Tormen 2001;
Seljak \& Warren, 2004; Reed \etal 2009; Pillepich, Porciani \& Hahn
2010; Tinker \etal 2010), the galaxy-halo connection then provides an
important avenue to investigate the density and velocity fields in the
universe from the observed galaxy distribution.

An empirical way to establish the galaxy-halo connection is to use
galaxy groups, provided that they are defined as sets of galaxies
that reside in the same dark matter halo. Recently, Yang et al.
(2005; 2007) have developed a halo-based group finder that is
optimized for grouping galaxies residing in the same dark matter
halo.  Using mock galaxy redshift surveys constructed from the
conditional luminosity function model (e.g. Yang, Mo \& van den
Bosch 2003) and semi-analytical models (Kang et al, 2005), it is
found that this group finder is very successful in associating
galaxies with their common dark matter haloes. The group finder
performs reliably for poor systems, including isolated galaxies in
small mass haloes, making it ideally suited for the study of the
relation between galaxies and dark matter haloes over a wide range
of halo masses (see also Weinmann \etal 2006 and Wang \etal 2008).
These galaxy groups, and the dark haloes they represent, together
with the relationship between dark haloes and the mass density
field predicted by the current CDM model, offer an unprecedented
opportunity to reconstruct the cosmic density and velocity fields
in the cosmic volume within which the galaxies are observed.

A powerful reconstruction method will also make it possible to
classify the morphology of the large scale structure. Most methods
aimed at describing the morphology of the large scale structure are
either based directly on the galaxy distribution (e.g. Hoyle \&
Vogeley 2002; Romano-D\'iaz \& van de Weygaert 2007; Sousbie et
al. 2011) or on the smoothed density field (e.g., Park et al. 2005;
Arag{\'o}n-Calvo et al. 2007). A particularly interesting
classification is based on the tidal field tensor, which is simply the
Hessian of the gravitational potential (e.g. Hahn et al. 2007a,b;
Forero-Romero et al. 2009; Arag{\'o}n-Calvo et al. 2010a,b).  Since
many properties of dark matter halos are found to be correlated with
the large-scale tidal field (e.g. Wang et al.  2011), such a
classification is particularly interesting for investigating how
galaxy properties are affected by their large-scale
environment. Hence, a reconstruction method that yields an accurate
estimate of the tidal field will open an important avenue for the
study of galaxy formation and evolution.

In an earlier paper (Wang \etal 2009; hereafter W09), we developed
a method to reconstruct the cosmic density, velocity and tidal
fields starting from the distribution of (massive) galaxy groups
(i.e., dark matter haloes), which we tested against $N$-body
simulations but did not apply to real data. The method to
reconstruct the {\it density} field partitions the volume in
domains associated with each individual group (i.e., dark matter
halo), and models the mass distribution in each domain using the
cross-correlation function between dark matter haloes and the mass
distribution within their domain obtained from $N$-body
simulations. The redshift distortions of the groups (domains) are
corrected iteratively by reconstructing the velocity field using
linear theory, in which the density field is traced by the most
massive groups, as described below.  This method is very different
from previous methods that have been used in earlier
investigations (e.g. Fisher et al. 1995; Zaroubi et al. 1995;
Erdo{\u g}du et al.  2004). In these studies the galaxy
distribution is usually smoothed heavily and normalized to
represent the cosmic density field on large scales. In the Wiener
reconstruction method, adopted in many of these earlier
investigations, the mass density at a given point is assumed to be
a linear combination of the observed galaxy density field values
at different points so that the reconstructed field has the
minimum mean square error. Therefore the results are expected to
be valid only on linear scales. Since our method is based on dark
matter haloes, the reconstruction is expected to be more accurate
both on small scales, where mass is strongly correlated with dark
matter halos (see also Kitaura et al. 2010; Jasche \& Kitaura 2010
for other methods to reconstruct the non-linear density field),
and on large scales, where the halo bias is well understood.

For the reconstruction of the velocity and tidal fields, W09 used
a slightly different method. Since the velocity field is mainly
dominated by the mass distribution on large scales, it can be
reconstructed simply from the distribution of the massive groups
(haloes), without having to account for their cross-correlation
with the dark matter in their domains (as we did for the
reconstruction of the density field). Using linear theory and a
smoothed density field based on the distribution of massive
haloes, W09 were able to accurately reconstruct the velocity and
tidal fields on large scales.  The velocity field thus obtained
was used in correcting the reconstructed density field for
redshift distortions.  This method for reconstructing the cosmic
velocity field is different from those used in earlier
investigations, which always used the galaxy distribution directly
(e.g. Kaiser et al. 1991; Willick \& Strauss 1998; Hudson et al.
2004), and therefore suffer from uncertainties in the complicated
relation between galaxies and the underlying mass distribution
(e.g., the fact that galaxy bias depends on luminosity and color).
The advantage of our method is that it is based on dark matter
halos, as represented by galaxy groups, so that the bias of the
distribution of different galaxies relative to the underlying
density field is automatically taken into account by their
connections to dark matter halos, whose bias is well understood.

The reconstruction method presented in W09 was recently used by
Munoz-Cuartas et al. (2011) to reconstruct the density field for
the survey volume of the SDSS DR4 (Adelman-McCarthy et al. 2006),
using the SDSS galaxy group catalogue of Yang \etal (2007).
However, we caution that Munoz-Cuartas et al. chose not to apply
the corrections for redshift distortion; hence their reconstructed
density field is in redshift space. In addition, they computed the
tidal deformation tensor, but did not properly account for
boundary effects of the SDSS volume, which, as we will see in this
paper, are very substantial. In this paper, we use the method of W09
to reconstruct the cosmic velocity field and the associated tidal
field in the SDSS survey volume. Our analysis differs
from that of Munoz-Cuartas et al. (2011) in that (i) we mainly
focus on the reconstruction of velocity and tidal fields, (ii) we
use the much larger SDSS DR7 (Abazajian et al. 2009) and its
corresponding galaxy group catalogue, (iii) we correct the
locations of the groups for redshift distortions, and (iv) we
carefully investigate the impact of boundary effects on the
accuracy of the reconstructed velocity and tidal fields. To that
extent we use detailed mock catalogues to test the reliability of
our method against uncertainties arising from redshift
distortions, survey boundaries, and false identifications of
groups by our group finder. We also present a new, alternative
method to reconstruct the density field, which does not require
$N$-body simulations to obtain the halo-matter
cross-correlation in halo domains. Rather, the method uses the
reconstructed velocity field and the Zel'dovich (1970)
approximation to infer the density field across cosmic
times. Note that although this method is much simpler, and less
time-consuming, it only yields an estimate of the smoothed density
field (i.e., it is unable to resolve highly non-linear regions).
Hence, this method should be considered complementary to the
method presented in W09, rather than as a true alternative.

This paper is organized as follows. In Section 2 we present the
galaxy and group catalogues used in this paper. Section 3
describes our reconstruction method, which is tested against mock
galaxy catalogues based on high resolution $N$-body simulations in
Section 4.  The reconstruction results based on the SDSS catalog
are presented in Section 5, and we summarize and discuss our
findings in Section 6.  Throughout this paper we adopt a WMAP5
cosmology (Dunkley et al.  2009): the density parameter
$\Omega_{\rm m}=0.258$; the cosmological constant
$\Omega_\Lambda=0.742$; the baryon density parameter
$\Omega_b=0.044$; the Hubble constant $h= 0.72$; and the linear
{\it RMS} density fluctuation in a sphere of an $8\mpc$ radius,
$\sigma_8$, equals 0.8.

\section{The SDSS catalogue}
\label{sec_sdss}

\begin{figure*}
\epsfig{file=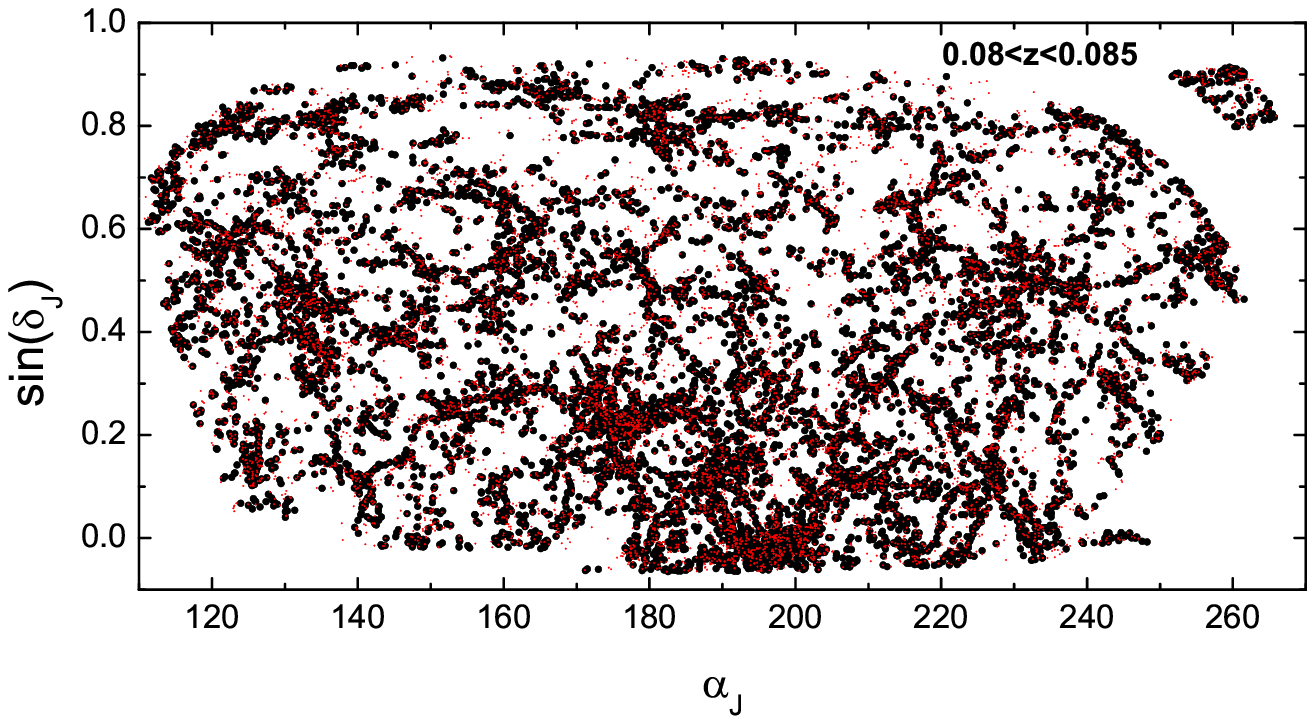} \caption{The distribution of the SDSS
groups with $\log(M_h/\msun)\ge 12$ (black dots), in comparison to
the distribution of galaxies that are assigned to halos with
smaller masses (red dots), in a specific redshift slice of
$0.08\leq z\leq0.085$.} \label{fig_ss}
\end{figure*}

The galaxy sample used here is constructed from the New York
University Value-Added Galaxy Catalogue (NYU-VAGC; Blanton et al.
2005), which is based on SDSS DR7 but includes a set of improved
reductions over the original pipeline. We use all galaxies in the
Main Galaxy Sample with extinction-corrected apparent magnitudes
brighter than $r=17.72$, with redshifts in the range $0.01 \leq z
\leq 0.20$, and with redshift completeness ${\cal C}_z > 0.7$. The
catalogue contains a total of $639,359$ galaxies with a sky
coverage of $7,748$ square degrees. For each galaxy we compute its
stellar mass from the $g$ and $r$-band magnitudes using the
method of Bell et al. (2003). DR7 covers two sky regions: a larger
region in the Northern Galactic Cap (NGC) and a smaller region in
the Southern Galactic Cap (SGC). As we demonstrate later, survey
boundaries can significantly impact the accuracy of our
reconstruction, and we therefore only use the more contiguous NGC
region (see Fig.~\ref{fig_ss} for a view of the sky coverage
used), which contains $584,473$ galaxies with a sky coverage of
$7,047$ square degrees.

Galaxy groups are selected using the adaptive halo-based group
finder developed by Yang et al. (2005). The application of this
group finder to the SDSS DR4 is described in detail in Yang et al.
(2007; hereafter Y07). The application to SDSS DR7 is exactly the
same, except that the sky coverage is significantly larger and we
adopt the WMAP5 cosmology, rather than the WMAP3 cosmology used by
Y07. The geometry of the SDSS used for the group catalogue is
defined as the region on the sky that satisfies the redshift
completeness criterion. As described in detail in Y07, our group
finder takes account of the survey edges in the SDSS volume by
estimating the fraction, $f_{\rm edge}$, of each group's volume
that falls inside of the survey volume. Group luminosities and
masses are then corrected for this fraction, and groups with
$f_{\rm edge} < 0.6$ are excluded, which removes only $1.6\%$ of
all groups.

As described in Y07, the majority of the groups in the catalogue
have two estimates of their dark matter halo masses: one based on
the ranking of the total characteristic luminosities of groups,
and the other based on the ranking of the total characteristic
stellar masses, both determined from group member galaxies more
luminous than $M_r - 5 \log h=-19.5$.  As shown in Y07, both halo
masses agree very well with each other, with an average scatter
that decreases from $\sim0.1$ dex at the low mass end to
$\sim0.05$ at the massive end.  In this paper we adopt the halo
masses based on the characteristic luminosity ranking. The
luminosity-based group masses are available for a total of
$355,482$ groups in our sample, which host a total of $502,651$
galaxies. The mass assignment based on characteristic luminosity
ranking is complete to $z\sim 0.12$ for groups with halo masses
$M_h\ga 10^{12}h^{-1}{\rm M}_\odot$, and to $z\sim 0.14$ for
$M_h\ga 10^{12.5}h^{-1}{\rm M}_\odot$ (see Y07). In our analysis,
we will use all groups with $M_h\ge M_{\rm th}=10^{12}h^{-1}{\rm
M}_\odot$ and so we restrict our reconstruction to the nearby
volume covering the redshift range $0.01 \leq z \leq 0.12$, which
we call the survey volume. Within this survey volume, the total
number of groups above our mass limit (i.e., with $M_h \geq M_{\rm
th}$) is $12,1922$. Fig.~\ref{fig_ss} shows the distribution of
these groups (black dots) as well as the distribution of galaxies
that are assigned to halos with smaller masses (hereafter `field
galaxies', red dots), in a specific redshift slice centered on
$z\sim 0.08$, chosen to show the SDSS `Great Wall' (Gott \etal
2005). As one sees clearly, the distribution of the field galaxies
closely follows the large-scale structure delineated by the more
massive groups.

To perform the reconstruction, we transform the redshift distances
and J2000.0 coordinates for each group into the following comoving
Cartesian coordinates:
\begin{eqnarray}
X&=&r(z)\cos{\delta_{\rm J}}\cos{\alpha_{\rm J}};\nonumber\\
Y&=&r(z)\cos{\delta_{\rm J}}\sin{\alpha_{\rm J}};\nonumber\\
Z&=&r(z)\sin{\delta_{\rm J}}
\label{eq_coor}\,.
\end{eqnarray}
Here $\alpha_{\rm J}$ and $\delta_{\rm J}$ refer to the J2000.0
right ascension and declination, respectively, and $r(z)$ is the
comoving distance at redshift $z$. For the WMAP5 cosmology adopted
here, the ranges for $X$, $Y$ and $Z$ axes of the survey volume
are $(-351.0, -1.5)$, $(-329.7, 302.0)$ and $(-22.7,330.3)$(in
units of $\mpc$), so the maximal scale along the three axes are
about $350$, $632$, and $353\mpc$, respectively.

\section{The reconstruction method}
\label{sec_method}

We now describe our method to reconstruct the velocity and tidal
fields for the SDSS survey volume. Our method closely follows that
of W09, but with a few small modifications that are required in
order to properly account for the complicated geometry of the SDSS
survey volume.

\subsection{Velocity field}
\label{sec_vf}

In the linear regime, the peculiar velocity can be derived as
\begin{equation}
{\bf v}=-\frac{1}{4\pi G \bar{\rho}a}\frac{\dot{D}}{D}\nabla\phi\,,
\label{eq_vphi}
\end{equation}
where $G$, $a$, $\bar{\rho}$ and $D$ are the gravitational
constant, the scale factor of the universe, the cosmic mean
density and the linear growth rate, respectively.  The quantity
$\phi$ is the peculiar gravitational potential and can be
calculated from the density perturbation ($\delta$) through the
Poisson equation:
\begin{equation}
\nabla^2\phi=4\pi G\bar{\rho}a^2\delta\label{eq_phi}\,.
\end{equation}
Combining these two equations, and working in Fourier space, we
have
\begin{equation}
\textbf{v}(\textbf{k})=H a f(\Omega) \frac{i\textbf{k}}{k^2}\delta
({\bf k})\,, \label{eq_vk}
\end{equation}
where $\textbf{v}(\textbf{k})$ and $\delta ({\bf k})$ are the
Fourier transforms of ${\bf v}$ and $\delta$, respectively, $H$ is
the Hubble constant, and $f(\Omega)=d\ln D/d\ln a\simeq\Omega_{\rm
  m}^{0.6}+{1\over 70}\Omega_{\Lambda}(1+\Omega_{\rm m}/2)$
(e.g. Lahav et al. 1991).

As described and demonstrated in W09, one can compute the velocity
field using only the density field represented by dark matter
haloes above a given mass threshold, $M_{\rm th}$. The predicted
velocity, ${\bf v}_{\rm h}({\bf x})$, based on this halo
distribution is tightly correlated with, and directly proportional
to, the real velocity, ${\bf v}({\bf x})$. In particular, we can write
that (see Colombi, Chodorowski, \& Teyssier 2007)
\begin{equation}
\textbf{v}_{\rm h}({\bf k})=H a f(\Omega) \frac{i{\bf
k}}{k^2}\delta_{\rm h}({\bf k})=b_{\rm hm}Haf(\Omega) \frac{i{\bf
k}}{k^2}\delta({\bf k}) =b_{\rm hm} {\bf v}({\bf k})\,,
\label{eq_vhk}
\end{equation}
where $\delta_{\rm h}({\bf k})$ is the Fourier transform of the
mass density contrast represented by the haloes with mass $M_{\rm
h} \geq M_{\rm th}$, hereafter $\delta_{\rm h}({\bf x})$. As
explicitly shown in W09, $b_{\rm hm}$ is the average bias of the
haloes with mass $M_{\rm h} \geq M_{\rm th}$, and therefore the
bias of the density field, $\delta_{\rm h}$, is given by
\begin{equation}\label{eq:bhm}
b_{\rm hm} = {\int_{M_{\rm th}}^{\infty} M \, b_{\rm h}(M) \, n(M)
\, {\rm d}M \over \int_{M_{\rm th}}^{\infty} M \, n(M) \, {\rm
d}M}\,,
\end{equation}
where $n(M)$ and $b_{\rm h}(M)$ are the halo mass function and the
halo bias function, respectively. This bias factor can also be
obtained through a comparison between the predicted velocities
based on dark matter particles and haloes, as was done in W09.
Thus, the velocity field can be reconstructed using only the
population of haloes above some mass threshold. This is fortunate,
since it means that we can use our group catalogue, which represents
massive dark matter haloes, to accurately reconstruct the cosmic
velocity field.

In order to reconstruct the velocity field in the SDSS survey
volume we proceed as follows. We first embed the survey volume in
a periodic, cubic box of $726\mpc$ on a side (In the following,
this will be referred to as the survey box to distinguish it from
the survey volume and the simulation box to be defined below). The
linear size of the survey box is chosen to be about $100\mpc$
larger than the maximal scale of the survey volume among the three
axes. We divide the box into $1024^3$ grid cells (which are $\sim
0.7\mpc$ on a side), and sort them into two types: survey grid
cells and non-survey grid cells, depending on whether or not the
center of the grid cell in question is located inside the survey
volume. We assign the mass of each group (with mass $\ge M_{\rm
th}$) on the survey grid according to its redshift-space
coordinates. This is done by assuming that the halo mass is
distributed homogeneously within a radius of $R_{200}/2$, where
$R_{200}$ is the virial radius of the halo. Since we will smooth
the density field on relatively large scales, our results are very
insensitive to how exactly we distribute the halo mass; for
example, we have verified that adopting a Dirac delta function,
rather than a top-hat sphere, yields results that are
indistinguishable. Non-survey grid cells are assigned a density
equal to the average mass density of the groups (with $M_{\rm h}
\geq M_{\rm th}$) in the survey volume (hereafter $\bar{\rho}_{\rm
h}$) so that the entire survey box has the same mean density as
the actual survey volume. Next we compute the overdensity field of
our groups (in redshift space), defined by
\begin{equation}
\delta_{{\rm h},i} = {\rho_{{\rm h},i} - \bar{\rho}_{\rm h} \over
\bar{\rho}_{\rm h}}
\end{equation}
where the index $i$ refers to the grid cell in question.

To correct for the redshift distortions due to the peculiar
velocities of the groups, we follow exactly the procedure
developed by W09. We first smooth the density field using a
Gaussian smoothing kernel with a mass scale of $\log(M_{\rm
s}/\msun)=14.75$, which corresponds to a Gaussian kernel size of
\begin{equation}
R_s\equiv{1 \over \sqrt{2\pi}}\,\left({M_{\rm s}\over\bar{\rho}}
\right)^{1/3}=7.93 \mpc
\end{equation}
for the WMAP5 cosmology adopted here\footnote{As tested and
described in W09, this is the optimal smoothing scale for the
redshift distortion corrections.}. Next we Fast Fourier Transform
(FFT) the smoothed overdensity field of our groups, and use
Eq.~(\ref{eq_vhk}) to compute $\textbf{v}({\bf k}) =
\textbf{v}_{\rm h}({\bf k})/b_{\rm hm}$, where $b_{\rm hm}$ is
computed using Eq.~(\ref{eq:bhm}). Fourier transforming
$\textbf{v}({\bf k})$ then yields the velocity field, which we use
to compute the peculiar velocity-corrected, cosmological redshift
of each group according to
\begin{equation}
z_{\rm corr} = {z_{\rm obs} - (v_{\rm los}/c) \over 1 + (v_{\rm
los}/c)}\,,
\end{equation}
where $v_{\rm los}$ is the line-of-sight component of the peculiar
velocity and $z_{\rm obs}$ is the observed redshift of the group
(i.e., the luminosity weighted average redshift of the group
members). Since the velocity field is computed using the redshift
space distribution of the groups, this method needs to be iterated
until convergence is achieved. Detailed tests with mock catalogs
suggest that two iterations are generally sufficient. Note that
the relatively large smoothing scale is adopted to suppress
non-linear velocities that cannot be predicted accurately with our
linear model. Thus, our method only corrects redshift distortions
in the linear and quasi-linear regime.  The choice of a large
smoothing scale prevents the iterations from getting trapped in a
local minimum. However, the correction is not sufficient in the
boundary regions, as we will show below.

Now that we have a sample of groups with corrected positions, we
can use the procedure described above to assign group masses on
the survey grids to obtain $\delta_{\rm h}$ in `real space'. In
order to obtain the velocity field on different scales, we smooth
the density field with several choices of $M_{\rm s}$. The values
of ${\bf v}_{\rm h}/b_{\rm hm}$ on all the grid points then
represent the predicted peculiar velocity field.

\subsection{Gravitational tidal field}
\label{sec_tf}

We describe the large-scale tidal field through the tidal tensor,
${\cal T}_{ij}$, defined as
\begin{equation}\label{eq_tij}
{\cal T}_{ij}=\partial_i\partial_j\phi\,,
\end{equation}
where $\phi$ is the peculiar gravitational potential which can be
calculated from the mass density field using the Poisson equation
[Eq.~(\ref{eq_phi})].  Since we want to derive the tidal field
using only haloes (galaxy groups) with mass $M_{\rm h} \geq M_{\rm
th}$, we write the corresponding gravitational potential,
$\phi_h$, as
\begin{equation}\label{eq_psh}
\nabla^2\phi_{\rm h}
 =4\pi G\bar{\rho}_{\rm h}a^2\delta_{\rm h}
 =4\pi G\bar{\rho}a^2\, (b_{\rm hm}\delta) \,
\left(\frac{\bar{\rho}_{\rm h}}{\bar{\rho}}\right)
 = b_{\rm hm}\frac{\bar{\rho}_{\rm h}}{\bar{\rho}}\nabla^2\phi\,,
\end{equation}
where $\delta_{\rm h}$ is again the mass density distribution of
groups with $M_{\rm h} \geq M_{\rm th}$ and ${\bar\rho}_{\rm h}$
is the corresponding mean mass density. We thus can derive $\phi$
and ${\cal T}_{ij}$ using galaxy groups. It is easy to see that
the bias factor, $b_{\rm hm}$, in the above equation is the same
as that in Eq.~(\ref{eq_vhk}). The value of $\delta_{\rm h}$ can
be obtained using exactly the same method as described in
Section~\ref{sec_vf}, and we smooth $\delta_{\rm h}$ with a
Gaussian smoothing kernel of a given mass $M_{\rm s}$. We then use
FFT to obtain the potential field, $\phi$, by solving the Poisson
equation, and derivative operators are applied (in Fourier space)
to derive the tidal tensors. Finally, the eigenvalues $T_1$, $T_2$
and $T_3$ ($T_1>T_2>T_3$) of the tidal tensor are obtained at each
grid point by diagonalizing the corresponding tidal tensor.

The tidal field impacts dark matter haloes, and their associated
baryonic material, with a net angular momentum, and plays an
important role in regulating the growth and shape of dark matter
haloes (e.g., White 1984; Hahn \etal 2007a,b; Wang et al. 2011).
Hence, the tidal field is an important description of the
environments in which haloes and galaxies reside. In particular,
the number of positive eigenvalues of the tidal tensor can be used
to define the morphologies of large scale structures (e.g. Hahn et
al. 2007a,b; Forero-Romero \etal 2009; Munoz-Cuartas \etal 2011).
If all of the three eigenvalues of a grid cell are positive, the
grid cell is classified as {\tt cluster}. Similarly, grid cells
with one or two negative eigenvalues are classified as {\tt
filament} or {\tt sheet}, respectively, while grid cells for which
all three eigenvalues are negative are classified as {\tt void}.
It is worthwhile to note that the tidal field is the second
derivative of the gravitational potential, while the velocity
field is the first derivative. The tidal field is therefore
proportional to the derivative of the velocity field and can be
used to characterize the convergence and divergence of the flow:
around a {\tt cluster} point, the three positive eigenvalues mean
that the flows along all directions are converging to that point;
around a {\tt filament} point the flow is convergent along two
directions but divergent along the third; around a {\tt sheet}
point the flow is convergent along one direction but divergent
along the other two; and around a {\tt void} point, the flow is
divergent in all directions.  All this meshes well with the
picture of the cosmic web formation in the current CDM model (e.g.
Bond et al. 1996), indicating that the classification of the
large-scale structure according to the signs of $T_1$, $T_2$ and
$T_3$ provides a useful description of the various patterns in the
cosmic density and velocity fields.

\section{Tests based on Mock Data}

Before applying our method to the actual SDSS data, we need to
gauge the reliability of our reconstruction. Although many tests
have already been presented in W09, here we focus specifically on
an application to the SDSS survey volume, carefully assessing the
impact of boundary effects associated with the complicated
geometry of the SDSS survey volume. To that extent we use a mock
SDSS group catalogue constructed from $N$-body simulations to
carry out a series of test to verify the reliability of our method
against uncertainties arising from redshift distortion, survey
boundaries, and false identifications of groups by our group
finder.

\subsection{The N-body simulation and Mock catalogs}

We use the ``Millennium Simulation'' (MS) carried out by the Virgo
Consortium (Springel et al. 2005). This simulation assumes a
spatially-flat $\Lambda$CDM model, with density parameter
$\Omega_{\rm m}=0.25$, baryon density parameter $\Omega_b=0.045$,
Hubble constant $h= 0.73$, and the linear {\it RMS} density
fluctuation in a sphere of an $8\mpc$ radius, $\sigma_8 = 0.9$.
Note that this cosmology is different from the WMAP5 cosmology
adopted here, but this should not be a problem as we use it only
to test our method. The CDM density field of this simulation was
traced by $2160^3$ particles, each having a mass of
$M_p\sim8.6\times10^8\msun$, in a cubic box of $500\mpc$
(comoving). The characteristic mass, $M_\ast$, defined to be the
mass scale at which the {\it RMS} of the linear density field is
equal to $1.686$, is $\log (M_\ast/\msun)\approx 12.8$.

Dark matter halos were identified using the standard
friends-of-friends algorithm (e.g. Davis et al. 1985) with a
linking length that is 0.2 times the mean inter-particle
separation. The mass of a halo, $M_{\rm h}$, is simply defined as
the sum of the masses of all the particles in the halo. These
halos are referred to as `real halos' in the following, to
distinguish them from the groups identified by the group finder
applied to the mock galaxy catalog described below.

Our construction of the mock galaxy and group catalogues here is
similar to that described in Y07. First, we stack
$3\times3\times3$ replicates of the simulation box and populate
the real haloes in the stacked boxes with galaxies of different
luminosities, using the conditional luminosity function (CLF)
model of van den Bosch et al. (2007). This CLF describes the halo
occupation statistics of SDSS galaxies, and accurately matches the
SDSS luminosity function and the clustering properties of SDSS
galaxies as function of their luminosity. Phase-space parameters
are assigned to the galaxies following the method described in
More \etal (2009). Briefly, the brightest galaxy (central) in each
halo is located at rest at the halo center, while the other
galaxies (satellites) are distributed spherically following an NFW
(Navarro, Frenk \& White 1997) number density distribution with
the concentration-mass relation of Macci\`o \etal (2007).  At the
assigned position of each satellite
 galaxy, one-dimensional velocities are drawn from a Gaussian with
a dispersion computed from the Jeans equation assuming isotropy.
Next we place a virtual observer at the center of the stacked
boxes and assign each galaxy ($\alpha_{\rm J}$, $\delta_{\rm J}$)
coordinates and a redshift, which is a combination of the galaxy's
cosmological redshift and its peculiar velocity along the
line-of-sight. Then we construct a mock galaxy catalogue by
mimicking the sky coverage of the SDSS DR7, taking detailed
account of the angular variations in the magnitude limits and
completeness of the survey (see Li et al. 2007 for details).
Finally, we apply the halo-based group finder of Yang et al.
(2005) to the mock galaxy catalog to obtain a mock catalogue of
groups with halo masses assigned according to their characteristic
luminosities, as described in Y07 (see Section~\ref{sec_sdss}).

Similar to the SDSS group catalogue, the mock group catalogue is
also complete to $z\sim 0.12$ for groups with halo masses $M_{\rm
h}\ga 10^{12}\msun$. We thus again adopt $M_{\rm th}=10^{12}\msun$
and restrict our reconstruction to the volume covering the
redshift range $0.01 \leq z \leq 0.12$. Note that this volume is
called survey volume, and the survey volume for the mock catalogue
is very similar to that of the real SDSS catalogue, so that the
uncertainties in reconstruction arising from the survey boundaries
are the same. The mock catalogue thus allows us to test the
accuracy of the reconstruction from the real data. For the MS
cosmology, Eq.~(\ref{eq:bhm}) gives $b_{\rm hm}=1.56$ for $M_{\rm
th} = 10^{12}\msun$, which is the value we will adopt for the mock
catalogue.

\subsection{Quantifying the SDSS Survey Volume}

Because the structure outside the survey volume is not modelled
accurately, the reconstruction is expected to be better in the
inner region of the survey volume than near its boundary.  To
quantify this boundary effect, we have to calculate the distance
of each survey grid cell to the boundary. Unfortunately, the
geometry of the survey volume is very complicated, and it is hard
to calculate, or even define, the distance to the boundary. Hence
we introduce a parameter, the filling factor $F$, to characterize
the closeness of a survey grid cell to the boundary. For each survey
grid cell, $k$, the filling factor $F$ is defined as the fraction
of grid cell centers in a spherical volume of radius $R_F$
centered on $k$, that are located within the survey volume. Hence,
$F$ is expected to be much less than unity for a grid cell located
close to the boundary, while $F \simeq 1$ for a grid cell that is
located more than a distance $R_F$ from any survey boundary. Thus,
the value of $F$ can be used to quantify the closeness of a survey
grid cell to the survey boundary. What remains is to specify the
radius $R_F$. If $R_F \rightarrow \infty$ then $F \rightarrow 0$
for all $k$, while $F \rightarrow 1$ for $R_F \rightarrow 0$. We
adopt $R_F = 80 \mpc$, for which we obtain a useful dynamic range
in values of $F$: for the SDSS DR7 survey
volume, 25.4\% of the survey grid cells have $F\ge0.9$, 38.9\%
have $F\ge0.8$, and 66.4\% have $F\ge0.6$.
We note that our main results are insensitive to
changes in $R_F$ of a factor of two.

\subsection{Testing the Reconstruction of the Velocity Field}

\begin{figure*}
\epsfig{file=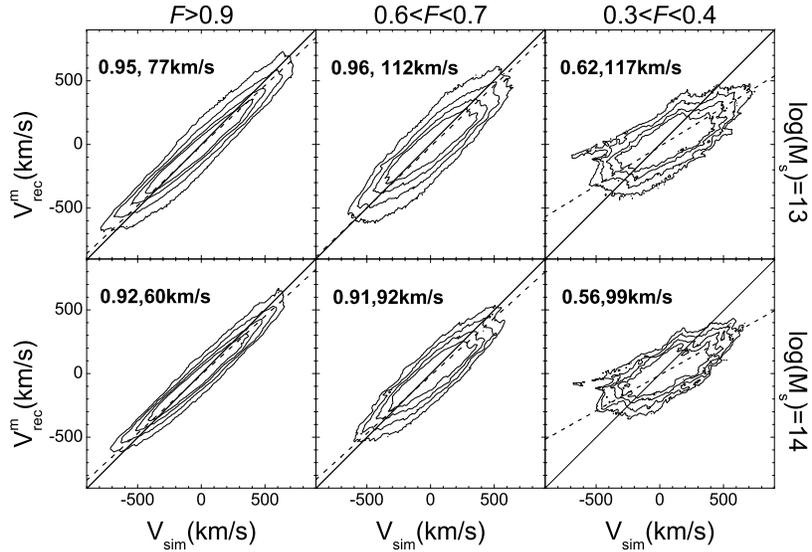}
\caption{The $Y$-component of the predicted velocity
based on the mock group catalogue versus the
corresponding velocity obtained from the simulation.
Results are shown for grid cells
within three different ranges of filling factors, $F$ (see text for
definition), and with two smoothing mass scales as indicated in
the figure. The four contours in each panel encompass 67\%, 80\%,
90\% and 95\% of the grid cells in a given range of $F$. The
velocity field and tidal field shown below are obtained based on
groups with $M_{\rm th}=10^{12}\msun$. The solid lines indicate
the unity slope relationship, while the dashed lines show the
best-fit linear relation of the correlation between the
reconstruction and the simulation. The first number in each panel is
the slope of the best-fit relation and the second number indicates
the scatter around the best-fit line.}
\label{fig_vmocky}
\end{figure*}

The upper panels of Fig. \ref{fig_vmocky} show the reconstructed
velocity (${\bf v}^{\rm m}_{\rm rec}$) obtained from the mock
catalogue using the method described in Section~\ref{sec_method}
versus the true velocity in the simulation (${\bf v}_{\rm
sim}$) (in contours), with smoothing mass scale $\log(M_{\rm
s}/\msun)=13$. Here, we only plot the comparison of the $Y$
component of the velocities, because its dynamic range is larger
than in the other two directions and because the results for all
the three directions are quite similar. Results are shown
separately for grid cells with $F\ge0.9$, $0.6\le F\le0.7$ and
$0.3\le F\le0.4$, as indicated. For grid cells with $F\ge0.9$
(i.e. which are far from any survey boundary), the reconstructed
velocity is tightly and linearly correlated with the true velocity
(the upper-left panel). We have performed a linear regression
and found a slope of 0.95. The best-fit line is shown in the
figure as the dashed line. The value of the slope and
the scatter around the best-fit relation are also
given in the panel. As $F$ decreases to $\sim 0.65$,
the slope of the correlation (0.96) does
not change significantly, but the tightness of the
correlation decreases. Finally, for grid cells
with $F \sim 0.35$, the best-fit slope of the correlation
(0.62) deviates significantly from unity. Based on these
results we conclude that our method can reliably reproduce the cosmic
velocity field in regions with $F \ge 0.6$, which applies to about
66\% of the SDSS survey volume, as mentioned above.

We then compare the results obtained above with those obtained with a
larger smoothing mass scale, $\log(M_{\rm s}/\msun)=14$. As $M_{\rm s}$
increases, the correlation becomes tighter while the slope and the
dynamic range of the velocity do not change significantly. The
slopes of the correlation for grid cells within
the three ranges of $F$ are now 0.92, 0.91 and 0.59
respectively (see the lower panels of Fig. \ref{fig_vmocky}).
This suggests that the velocity field is
produced primarily by relatively large-scale structures. Indeed,
for the CDM model considered here, the velocity field is dominated
by structure on scales where the effective power index of the
power spectrum is $\sim -1$ [see Chapter~6 in Mo, van den Bosch \&
White (2010; hereafter MBW10)], which corresponds to a mass scale
of $\sim 10^{15}\msun$.

\begin{figure*}
\epsfig{file=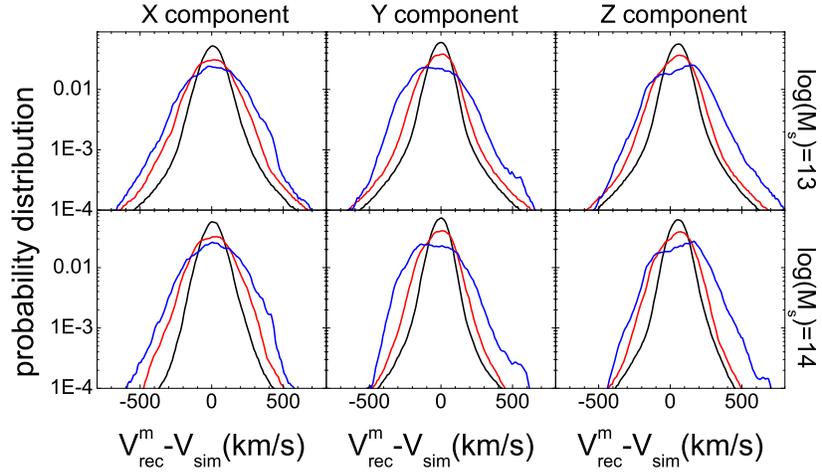,scale=1.0} \caption{The probability
distribution of the difference between the predicted and real
velocities along $X$, $Y$ and $Z$ axes (left, middle and right
panels respectively). Results are shown for two choices of
smoothing mass scales, as indicated. The black, red and blue lines
in each panel show the results for grid cells with $F\ge0.9$, $0.6\le
F\le0.7$ and $0.3\le F\le0.4$, respectively. }\label{fig_vpdf}
\end{figure*}

\begin{figure*}
\epsfig{file=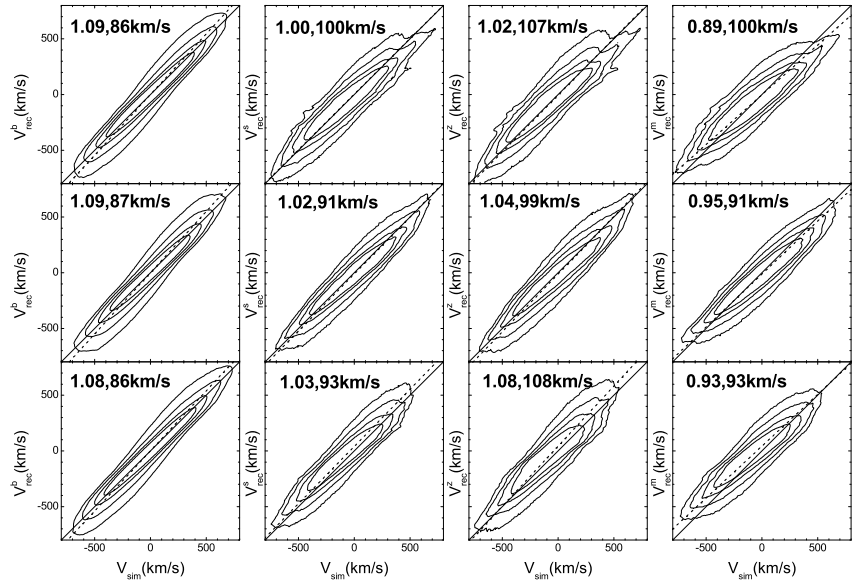} \caption{The predicted versus simulation
velocities based on different group catalogues. The top, middle
and bottom panels show the results along $X$, $Y$ and $Z$ axes,
respectively. The four contours in each panel encompass 67\%,
80\%, 90\% and 95\% of the grids. Results are only shown for
$\log(M_s/\msun)=13$. The first column panels show the predicted
velocity based on real halos distributed in real space and in the
periodic simulation box. The second column panels show the results
based on real halos distributed in real space and in the survey
volume. The third column panels show the results based on real
halos distributed in redshift space and in the survey volume. The
fourth column panels show the results based on the mock group
catalogue. Note that redshift distortion is corrected for the last
two cases. The dashed lines are the best linear fit,
and the numbers are the slopes of these best fit
relation and the scatter of the correlation relative
to the best fit.}
\label{fig_vdm}
\end{figure*}

Fig. \ref{fig_vpdf} shows the probability distribution of the
difference between the reconstructed and true velocities ($v^{\rm
m}_{\rm rec}-v_{\rm sim}$).  This confirms that the reconstruction
is better for larger $F$ and larger $M_{\rm s}$. For $F\ge0.6$,
the distributions for the $X$ and $Y$ velocity-components peak
nicely at zero, while for the $Z$-component the peaks are located
at $+50\kms$, indicating a slight bias in the reconstruction. In
fact, there is weak but systemic bias between $v^{\rm m}_{\rm
rec}$ and $v_{\rm sim}$ along all the three axes (see Fig.
\ref{fig_vmocky} and Fig. \ref{fig_vdm}), in the sense that the
absolute value of $v^{\rm m}_{\rm rec}$ is, on average, smaller
than that of $v_{\rm sim}$. These systematic deviations can be
caused by a number of factors. First of all, they may signal
inaccuracies associated with the method itself. For example, the
linear theory used to infer the velocity field from the gradient
of the peculiar potential [Eq.~(\ref{eq_vphi})] is expected to be
inaccurate for grid cells where nonlinear effects are not
negligible. Secondly, the problem may arise from the fact that the
survey volume is finite: velocity is generated by gravity, which
is a long range force. Hence, if the survey volume is too small,
the gravitational force due to the mass distribution outside the
survey volume may make a significant contribution which is not
included in our calculation. Thirdly, it may be that our
correction for redshift distortions turns out to be inaccurate.
And finally, systematic deviations in the reconstructed velocity
field may also arise from inaccuracies (i.e., false
identifications of groups) in our group finder.

In order to determine which, if any, of these factor(s) dominate,
we perform a series of tests. For all these tests we keep the
smoothing mass scale fixed at $\log(M_{s}/\msun)=13$. For our
first test we use real halos distributed in real space in the
periodic simulation box ($500\mpc$) to predict the velocity ($v^{\rm
b}_{\rm rec}$) using linear theory. In this case, the
reconstruction does not suffer from effects due to limited volume,
redshift distortions or our group finder. Hence, a comparison
between the reconstructed and true velocity fields in this case
tests the accuracy of our method in the absence of `observational'
effects. The first column of Fig.~\ref{fig_vdm} shows the
reconstructed velocity versus the true velocity along all three
axes. The corresponding slope and scatter of the correlation
are shown in each panel for comparison. As one can see, the
correlations are steeper than a slope of unity and the deviation
for all the three axes are similar, about 9\%. This
clear shows that our method is not perfect, likely because
of the use of linear theory in our model.
To check the effects due to the limited
survey volume, we use real halos distributed in real space, but
only those located within the {\it survey volume}, to predict the
velocity ($v^{\rm s}_{\rm rec}$). The resulting correlation
between $v^{\rm s}_{\rm rec}$ and $v_{\rm sim}$  for survey grid
cells with $F\ge0.6$ is shown in the second column of
Fig.~\ref{fig_vdm}. The $v^{\rm s}_{\rm rec}$ - $v_{\rm sim}$
correlations are flatter than the $v^{\rm b}_{\rm rec}$ - $v_{\rm
sim}$ correlations for all three velocity components.
On average the slope decreases by about 7\%, indicating that the
large-scale structure outside the survey volume indeed contributes
to the acceleration. Next we consider the effect due to redshift
distortion. Using the velocities of real halos, we generate a halo
sample in redshift space within the {\it survey volume}. We then
use the method described in Section \ref{sec_vf} to correct for
the redshift distortion and apply our reconstruction method to
estimate the velocity field ($v^{\rm z}_{\rm rec}$). In this case,
the correlation for cells with $F \ge 0.6$ (shown in the third
column) is similar to that between $v^{\rm s}_{\rm rec}$ and
$v_{\rm sim}$ (with an increase of 3\% in the slope),
suggesting that our method for correcting the redshift distortions
works well and does not introduce significant bias. Finally, in
the fourth column of Fig.~\ref{fig_vdm} we show the correlation
between $v^{\rm m}_{\rm rec}$ and $v_{\rm sim}$ (again only for
cells with $F \ge 0.6$), which includes also the effect of our
group finder. Here the correlation becomes flatter than the
$v^{\rm z}_{\rm rec}$ - $v_{\rm sim}$ correlation for all the
three velocity components, albeit by a small amount
(12\% in the slope). This indicates that the group
finder itself also introduces a weak bias.

The amplitudes of the deviation from a unity-slope relation
introduced by the inaccuracy of the method (9\%), the limited
volume effect ($-$7\%), and the group finder ($-$12\%) are roughly
comparable, and so all of them are important for producing the
final deviation. Because this bias is relatively small and because
it is extremely complicated to model accurately, we do not try to
correct for it in what follows. Note that the scatter in the
reconstructed velocity field (which is indicated in each panel) is
mainly due to the inaccuracy of the method. We have also checked
the distribution of the residual between the true velocity and
each of the four reconstructed velocities. For the first case,
where real halos distributed in real space in the periodic
simulation box are used, the distritions of the three velocity
components are identical. For the other three cases, where halos
are distributed in the survey volume, the distributions are
similar to those shown in Fig. 3 (see also Fig. 4). This suggests
that the differences between the three components in our mock
tests are mainly due to the survey geometry.

\subsection{Testing the Reconstruction of the Tidal Field}

\begin{figure*}
\epsfig{file=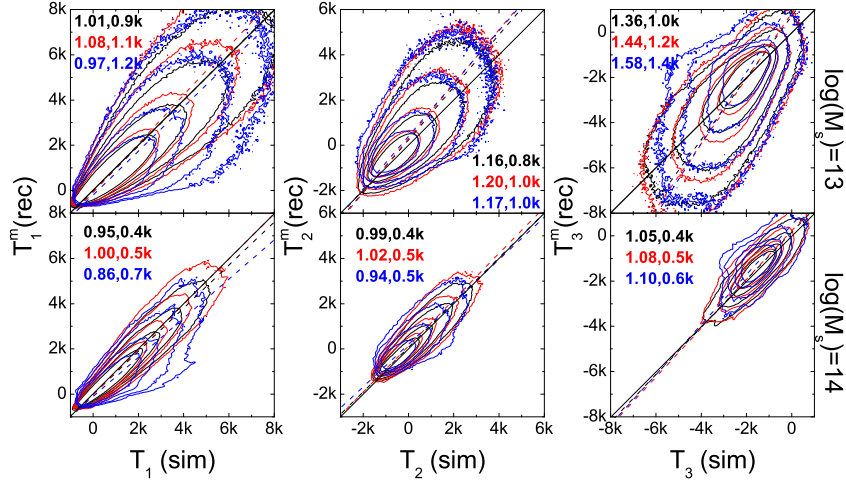}
\caption{The three eigenvalues of the predicted
tidal tensor from the mock group catalogue versus
these obtained from the simulation. Results are shown
for two choices of the smoothing mass scale, as indicated. Results
for grid cells with $F\ge0.9$, $0.6\le F\le 0.7$ and $0.3\le F\le0.4$
are shown in black, red and blue, respectively. The four contours
for each result encompass 67\%, 80\%, 90\% and 95\% of the grid
cells, respectively. The dashed lines are the best-fit, and the
numbers are the slopes of the best-fit and the scatter
relative to the best fit.}
\label{fig_tmock}
\end{figure*}

\begin{figure*}
\epsfig{file=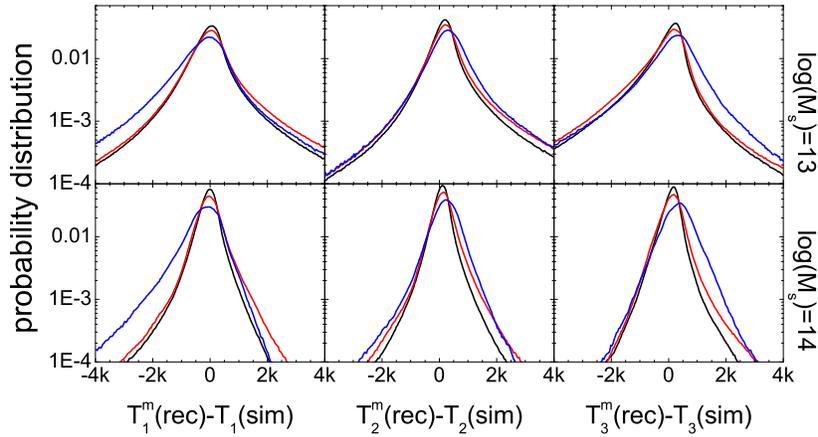} \caption{The probability distribution
of the difference between the predicted eigenvalues of the tidal
tensor and the eigenvalues obtained from simulation. The results
with two smoothing mass scales are shown, as indicated. The black,
red and blue lines in each panel show the results for grid cells
with $F\ge 0.9$, $0.6\le F\le0.7$ and $0.3\le F\le0.4$, respectively.
}\label{fig_tpdf}
\end{figure*}

We now move to the reconstruction of the cosmic tidal field.
Fig.~\ref{fig_tmock} shows the tidal field reconstructed from the
mock catalogue, $T^{\rm m}_i(\rm rec)$, versus the tidal field
calculated directly from the MS simulation, $T_i(\rm sim)$,
together with the best-fit linear relations and
the corresponding values of the slope and the scatter.
Fig.~\ref{fig_tpdf} shows the probability
distribution of $T^{\rm m}_i({\rm rec})-T_i(\rm sim)$ obtained for
different choices of $F$ and $M_{\rm s}$. Overall, the eigenvalues
of the reconstructed tidal tensor, $T^{\rm m}_i(\rm rec)$, are
strongly correlated with $T_i(\rm sim)$, albeit with a
considerable bias and a large scatter for small $M_{\rm s}$. The
bias is almost absent and the scatter becomes much smaller as we
adopt a larger smoothing scale, $\log(M_{\rm s}/\msun)=14$. Moreover
the reconstruction is better for grid cells with larger $F$,
similar to the velocity field. However, our results clearly show
that the effect of $F$ is not as important for the tidal field as
for the velocity field. The reason is that the tidal field, being
the derivative of the velocity field, is more dominated by smaller
scale structure. This is also evident from the fact that the
dynamic range of the reconstructed tidal field decreases
significantly when $\log(M_{\rm s}/\msun)$ increases.

\begin{figure*}
\epsfig{file=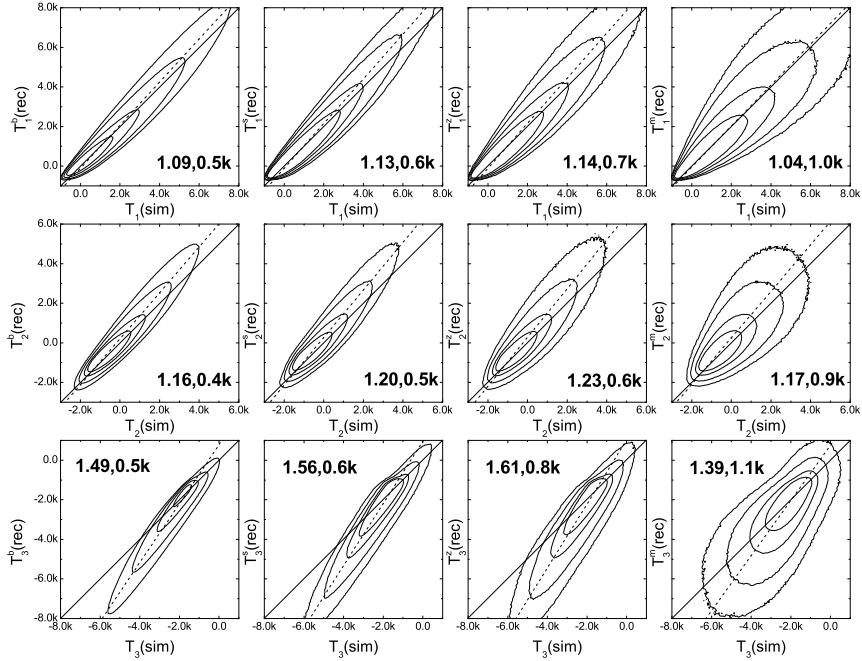} \caption{The same as Fig. \ref{fig_vdm}
but for the eigenvalues of the tidal tensor.} \label{fig_tdm}
\end{figure*}

We have also carried out a series of tests to investigate what
cause the scatter and bias in the tidal field reconstruction. As
for the velocity field, we use the following four cases:
(i) real halos in real space in the periodic simulation box;
(ii) real halos in real space in the survey volume;
(iii) real halos in redshift space and in the survey volume;
and (iv) mock group catalogues.  The
reconstructed tidal fields are denoted by $T^{\rm b}_i(\rm
{rec})$, $T^{\rm s}_i(\rm {rec})$, $T^{\rm z}_i(\rm {rec})$ and
$T^{\rm m}_i(\rm {rec})$, respectively.  The comparisons of these
reconstructions with the simulation results are shown in
Fig.~\ref{fig_tdm} for grid cells with $F\ge0.6$ (for the first
test, the results for all grid cells are shown.). For all these
tests we have adopted $\log(M_{\rm s}/\msun)=13$. Two things are
worth noting. One is that the large scatter in the $T^{\rm
m}_i({\rm rec})$-$T_i({\rm sim})$ correlation is mainly caused by
the inaccuracy of the method and the group finder, although other
two effects also contribute to it. The other is that the deviation
of the $T^{\rm m}_i({\rm rec})$-$T_i({\rm sim})$ correlation from
a slope of unity is inherited from the $T^{\rm b}_i({\rm
rec})$-$T_i({\rm sim})$ correlation, almost independent of the
effects due to redshift distortion, limited volume and group
finder. This deviation almost disappears if we adopt $\log(M_{\rm
s}/\msun)=14$ (see Fig. \ref{fig_tmock}), strongly suggesting that
it is caused by non-linear effects on small scales.

\section{Application to the Sloan Digital Sky Survey}

So far we have demonstrated that our method is able to reliably
reconstruct the velocity and tidal fields in regions with
$F\ge0.6$. In this section, we apply our method to the galaxy
groups selected from the SDSS DR7. As noted in Section
\ref{sec_sdss}, we use groups with $\log(M_h/\msun)\ge12$ to
perform the reconstruction in the survey volume ($0.01 \leq z \leq
0.12$). For the WMAP5 cosmology and $M_{\rm hm}=10^{12}\msun$ we
obtain $b_{\rm th}=1.69$ from Eq.~(\ref{eq:bhm}), which is the
value we adopted throughout. While we adopt a mass smoothing scale
of $\log(M_{\rm s}/\msun)=14.75$ for the correction for redshift
distortions (as described in Section~\ref{sec_vf}), the results
presented below have been obtained using a smoothing mass scale of
$\log(M_{\rm s}/\msun)=13$. Choosing the later scale to smooth our
finial results is a tradeoff between the accuracy and dynamic
range.

\subsection{The Tidal Field and the Classification
of the Large-Scale Structure}

\begin{figure*}
\epsfig{file=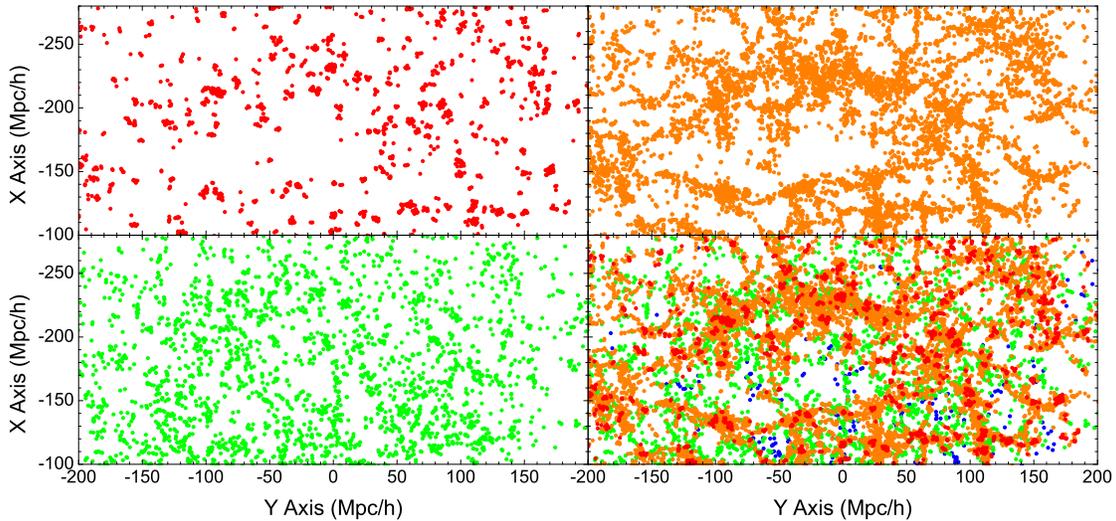} \caption{Classification of the large
scale structure in a slice $16\mpc$ thick enclosing the
SDSS Great Wall. The red dots are groups located at points
classified as cluster. The orange dots are the groups located at
points classified as filament. The green dots are groups located
at sheet points, while the blue dots in the lower right
panel are groups located at void points.
Both groups with masses $M_h\ge
M_{\rm th}$ and $M_h<M_{\rm th}$ are shown. Positions of these
groups are corrected for redshift distortion.} \label{fig_lss}
\end{figure*}

Let us first look at the classification of the large-scale
structure in the SDSS DR7 based on the reconstructed tidal tensor
(see Section~\ref{sec_tf}). Hahn et al. (2007a,b) found that using
a smoothing mass scale of $M_{\rm s}=2M_{\ast}$ yields good
agreement with the visual classification of the large-scale
structure. Since for the WMAP5 cosmology $\log
(M_\ast/\msun)\approx 12.5$, our adopted smoothing scale of
$M_{\rm s}=10^{13}\msun$ nicely satisfies that criterion.
Fig.~\ref{fig_lss} shows groups located in regions classified as
{\tt cluster} (red dots), {\tt filament} (orange dots) and {\tt
sheet} (green dots) in a $16\mpc$ thick slice enclosing the SDSS
Great Wall. Note that both groups with $M_h\ge M_{\rm th}$ and
$M_h<M_{\rm th}$ are shown here. The orange dots nicely delineate
the filamentary structure connecting the red dots, while green
dots are more diffused, forming `envelops' around the orange dots
(lower-right panel).  Such `hierarchical structure' nicely accords
with the CDM scenario of structure formation. {\tt Void} groups
are also shown, as blue dots in the lower-right panel, but since
groups in voids are rare, it is difficult to use the distribution
of groups to characterize a void region. We therefore show the
grid cells in {\tt void} in cyan in a plane with one cell thick
centered at ($-159$, $90$, $230$) (in units of $\mpc$), chosen to
represent a large underdense region. The surrounding grid cells in
{\tt cluster}, {\tt filament} and {\tt sheet} are shown in red,
orange and green respectively. For this particular region, the
size of the void is about $100\mpc$. We also compute the volume
filling fractions of the four different structures. The fractions
in the $F\ge 0.6$ region of the SDSS survey volume are about
1.9\%, 31.8\%, 53.2\% and 13.1\%, for {\tt cluster}, {\tt
filament}, {\tt sheet} and {\tt void}, respectively, in good
agreement with the results obtained from simulations (Hahn et al.
2007a; Forero-Romero et al. 2009) and from the SDSS galaxy
redshift survey (Jasche et al. 2010).

\subsection{The Velocity Field}

\begin{figure*}
\epsfig{file=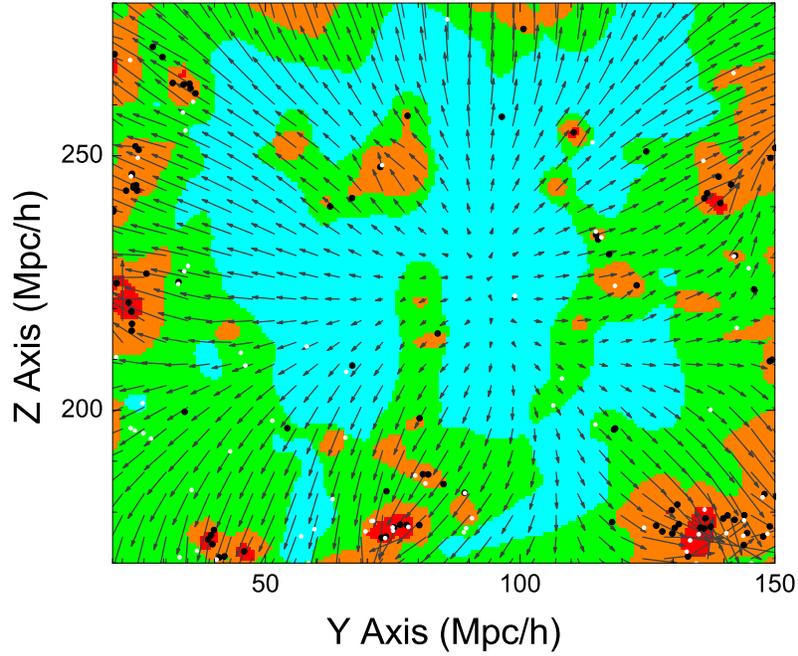} \caption{The velocity field in the $Y$
- $Z$ plane with $X=-159\mpc$ (grey vectors),  together with the
distribution of groups (dots) in a slice $4\mpc$ thick. The black
dots are groups with masses larger than $10^{12}\msun$, while
the white dots are the rest of the groups. Positions of the groups
are corrected for redshift distortion. The grid cells in
cluster, filament, sheet and void are show in red, orange, green
and cyan, respectively.} \label{fig_void}
\end{figure*}

\begin{figure*}
\epsfig{file=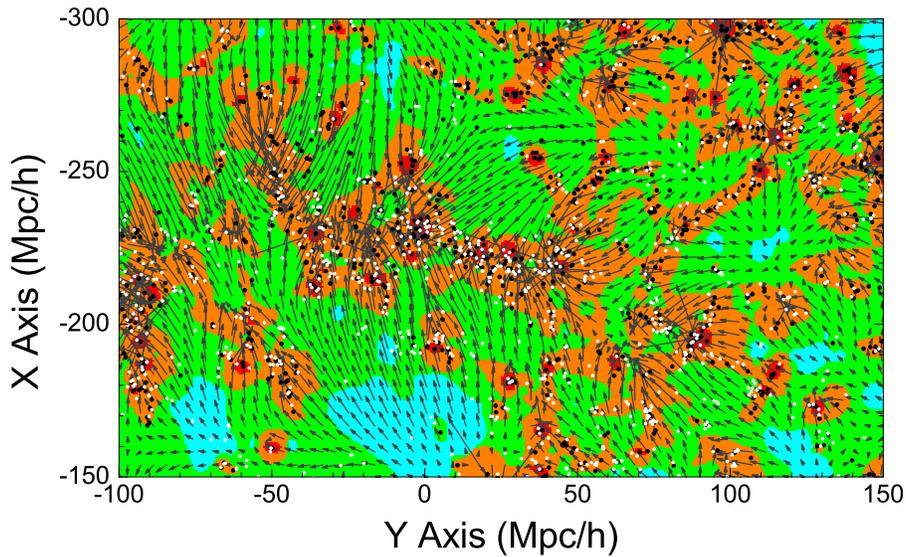} \caption{The velocity field in the $Y$ -
$X$ plane with $Z=8\mpc$ (gray vectors), together with the
distribution of groups (dots) in a slice  $4\mpc$ thick. The
symbols and color coding are the same as in Fig. \ref{fig_void}.
Positions of the groups are corrected for redshift distortion.
Note that the massive structure shown here is part of the SDSS
Great Wall. } \label{fig_gw}
\end{figure*}

Next let us look at the velocity fields in different regions. The
arrows in Fig.~\ref{fig_void} show the velocity vectors at the
grid points, with their lengths proportional to their magnitudes.
Note that the void region reveals a clearly divergent flow
emerging from its center, and with a magnitude that increases with
the distance from the center. The pattern is quite different in
high density regions, such as the SDSS Great Wall shown in
Fig.~\ref{fig_gw}. Here the flow is clearly converging towards the
central structure, and the velocity vectors have a tendency to be
perpendicular to the filamentary structure. This behavior can be
understood in terms of the Zel'dovich approximation (Zel'dovich
1970), which predicts that the velocity flow tends to be
perpendicular to the largest dimensions of pancakes and filaments.
To enhance the visual connection between the velocity field and
the large scale structures, we also plot in these figures the
distributions of both massive groups ($M_{\rm h} \ge M_{\rm th}$;
black dots) as well as low mass groups ($M_{\rm h} < M_{\rm th}$;
white dots) in slices $4\mpc$ thick, and also show the {\tt
cluster}, {\tt filament}, {\tt sheet} and {\tt void} regions in
red, orange, green and cyan, respectively.

\begin{figure*}
\epsfig{file=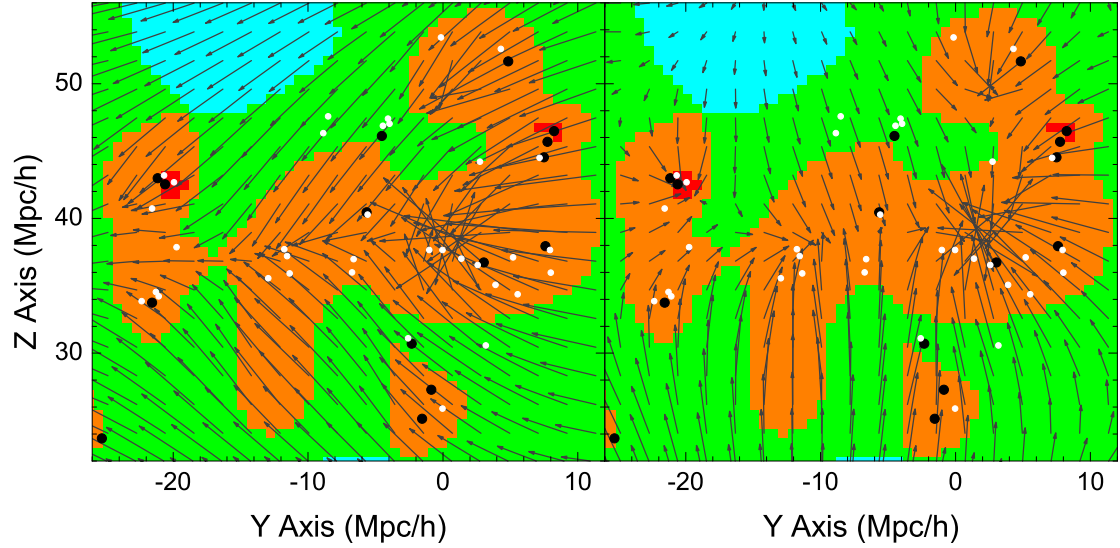,scale=1.} \caption{The velocity field
(left panel) and the velocity field relative to the center of mass
velocity (right panel) of a small structure in the $Y$- $Z$ plane
with X$= -159\mpc$ (gray vectors), together with the distribution
of groups (dots) in a slice $4\mpc$ thick. Positions of the
groups are corrected for redshift distortion. The symbols and
color coding are the same as in Fig. \ref{fig_void}. }
\label{fig_sstr}
\end{figure*}

Convergent flow are also predicted to be present around smaller
filamentary structures. We select a relatively small structure
($38\mpc \times 34\mpc$ comoving) and show its surrounding
velocity field in the left panel of Fig.~\ref{fig_sstr}, using the
same color coding as in Fig.~\ref{fig_gw}. This small structure
has apparent bulk motions towards negative $Y$ direction, which
may be induced by nearby massive structures and dominate over the
peculiar velocities produced by the more local mass distribution.
However, we still can see the signal of convergent flow. To
demonstrate this more clearly, we show the velocity field {\it
relative to the velocity of its center of mass} in the right
panel. The velocity flow clearly converges towards and is
perpendicular to the small filament. Interestingly, the flow seems
to converge predominantly towards the low mass groups with $M_{\rm
h} < M_{\rm th}$ (white dots), which are not used in our
derivation of the velocity or tidal fields. This suggests that the
large scale tidal field can compress masses into
filamentary/sheet-like structures, in accord with the Zel'dovich
approximation (Zel'dovich 1970). Note that the results shown in
Fig. \ref{fig_lss}, \ref{fig_void}, \ref{fig_gw} and
\ref{fig_sstr} are not restricted to $F\ge 0.6$ regions, although
most of the regions in the first three figures, and all of the
regions in the last figure, have $F\ge 0.6$.

\begin{figure*}
\epsfig{file=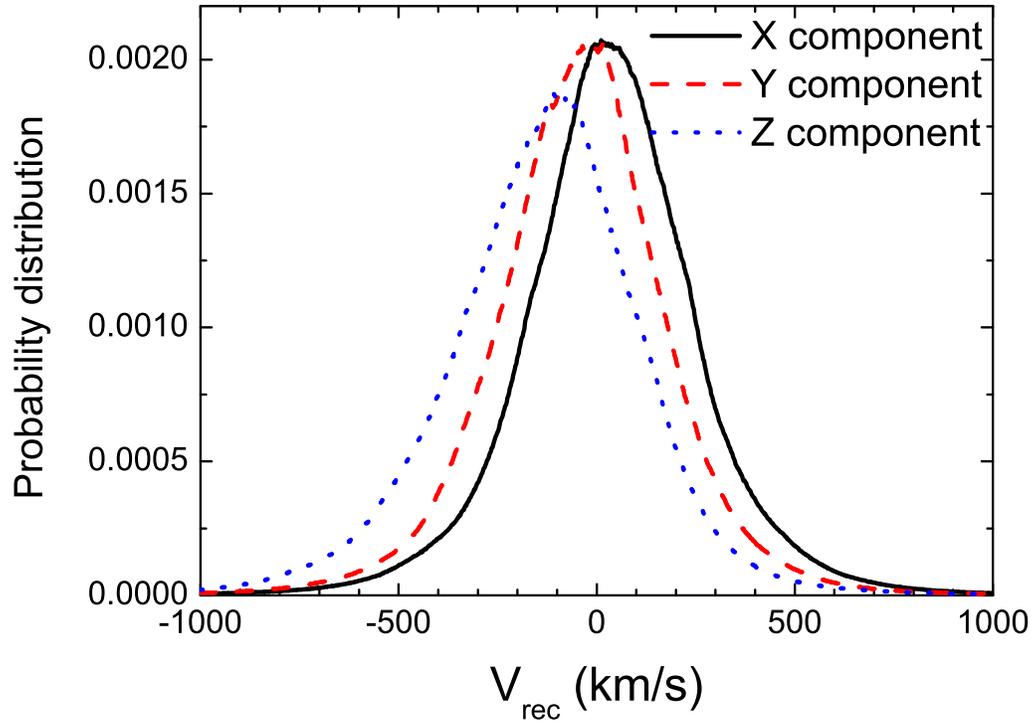,scale=1.3} \caption{The probability
distributions of the predicted velocities on grid cells with $F\ge0.6$
in $X$, $Y$ and $Z$ directions. The smoothing mass scale is
$10^{13}\msun$.} \label{fig_vdis}
\end{figure*}

Finally, in Fig.~\ref{fig_vdis} we show the probability
distribution of the three velocity components ($v_{\rm x}$,
$v_{\rm y}$ and $v_{\rm z}$) for grid cells with $F\ge0.6$. As
before, we have adopted a smoothing mass scale of $\log(M_{\rm
s}/\msun)=13$. Using $\log(M_{\rm s}/\msun)=14$ or 15 gives quite
similar results, although the dynamic range decreases slightly as
$\log(M_{\rm s}/\msun)$ increases. The distribution of each
velocity component is approximately Gaussian with extended wings,
consistent with the finding of Sheth \& Diaferio (2001). The
best-fitting Gaussian profiles for both the $X$ and $Y$ components
peak roughly at zero, with a dispersion of about $360\kms$, again
in good agreement with the prediction of the current $\Lambda$CDM
model (Sheth \& Diaferio 2001). However, the Gaussian profile for
the $Z$-component peaks at about $-117\kms$ and has dispersion of
$413\kms$. This suggests that a bulk flow is present over a volume
about 66\% of the survey volume (where $F\ge 0.6$), which is
equivalent to a sphere with a radius of $\sim 170\mpc$. Such a
bulk flow implies the existence of one or more massive structures
at small $Z$ (i.e. small $\delta_{\rm J}$) in the survey volume.
Such a large-scale inhomogeneity is indeed present, in the
form of the SDSS Great Wall, which is located near
$\delta_{\rm J}=4^{\circ}$, with a redshift of 0.08 (Gott et al.
2005; see also Fig.~\ref{fig_ss}).  It would be interesting to
investigate such large-scale flows further in comparison with
those given by other methods, such as measurements based on
distance indicators (e.g. Watkins et al. 2009; Feldman et al.
2010).

\subsection{A New Method to Reconstruct the Density Field}

As described in Section~\ref{sec_intro}, W09 presented a new
method to reconstruct the matter density field from the
distribution of (relatively) massive groups (i.e., dark matter
haloes). The method partitions the volume in domains associated
with each individual group, and models the mass distribution in
each domain using the cross-correlation function between dark
matter haloes and the mass distribution within their domain
obtained from $N$-body simulations. Here we present an alternative
method to reconstruct the density field, which is significantly
less elaborate in that it does not require $N$-body simulations to
characterize the halo-matter cross-correlation in halo domains.

Rather, our new method uses the Zel'dovich (1970) approximation to
displace mass elements using the velocity field obtained using the
method described in Section~\ref{sec_method} (see also Nusser et
al. 1991). According to the Zel'dovich approximation, the growth
of structure can be specified by the displacement of mass elements
from their initial positions, and the displacement,
$\textbf{r}-\textbf{r}_i$, of each mass element is proportional to
the gradient of the {\it initial} gravitational potential at the
initial position, $\textbf{r}_i$. Since the potential $\phi\propto
D(a)/a$ and ${\bf v}\propto\nabla\phi$ [equation (\ref{eq_vphi})],
we can also use the velocity field at redshift zero, instead of
the initial potential, to calculate the displacement:
\begin{equation}\label{eq_zel}
\textbf{r}=\textbf{r}_i+\frac{\textbf{v}_0(\textbf{r}_i)}
{H_0a_0f(\Omega_0)}\frac{D(a)}{D(a_0)}\,,
\end{equation}
[see \S~4.1.8 of MBW10] where $\textbf{v}_0(\textbf{r}_i)$ is the
velocity vector at redshift zero at the initial position of the
mass element.  Since the initial density perturbations are small,
one may use particles located on a uniform grid to sample the
initial density field. The above equation can then be used to
predict the positions of the particles at $a=1$ ($z=0$) (or at any
other redshift), thereby obtaining the density field.

\begin{figure*}
\epsfig{file=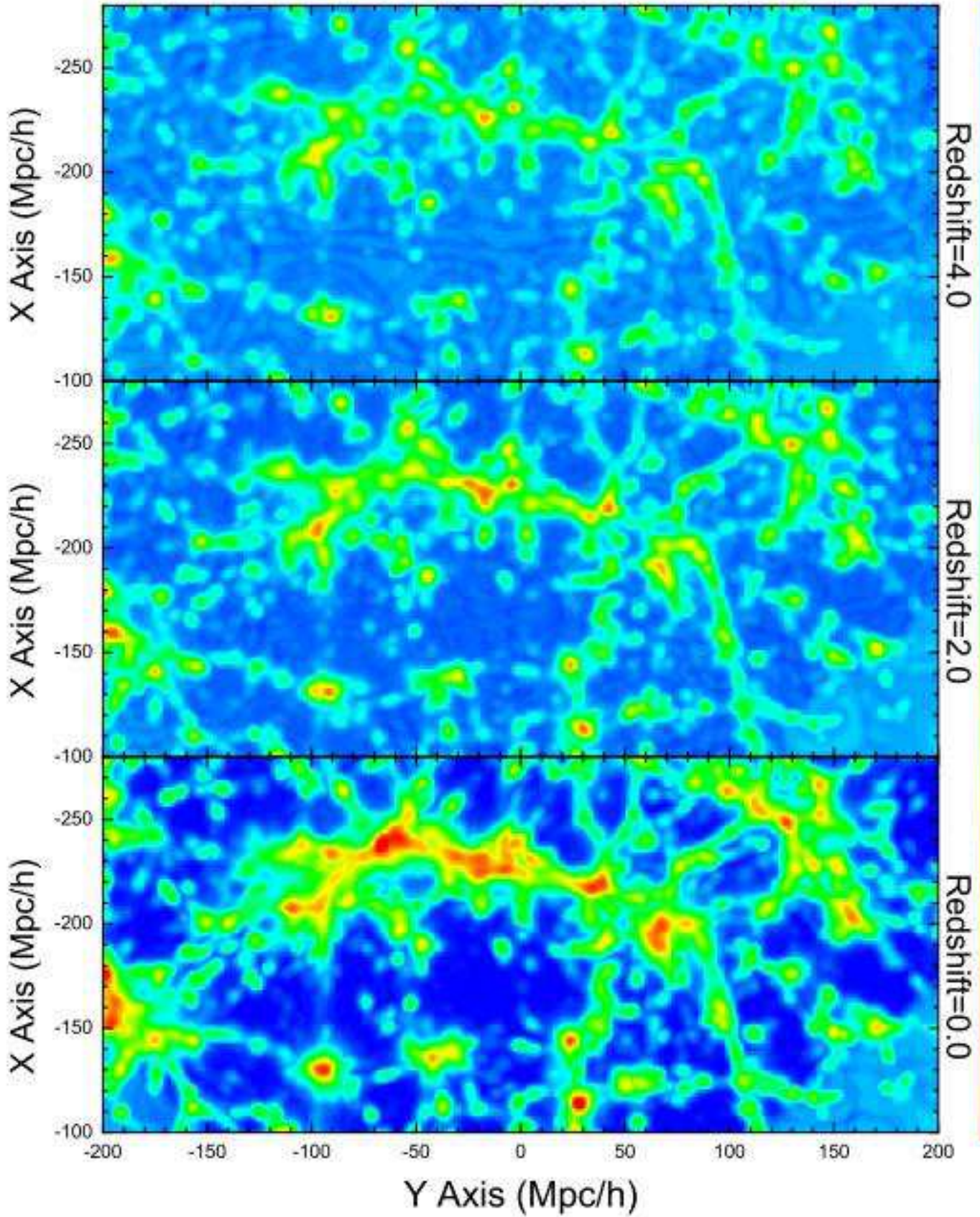} \caption{The evolution of the density
field around the SDSS Great Wall. The density field is produced
using Zel'dovich approximation and the predicted velocity field.
The color of each grid cell corresponds the logarithm of the
density field ($\log{\rho/{\bar\rho}}$), as indicated in the
figure.} \label{fig_zeld}
\end{figure*}

Here we apply Eq.~(\ref{eq_zel}) to our reconstructed velocity
field for the SDSS DR7 survey volume to predict the
corresponding large-scale density field. We start by generating a
sample of $512^3$ particles uniformly distributed in the survey
box, which we subsequently displace using the reconstructed
velocity field. The resulting density field at redshift of zero,
smoothed with a Gaussian kernel with a mass scale of $\log(M_{\rm
s}/\msun)=13$, is shown in the bottom panel of Fig.
\ref{fig_zeld}.  Comparing this reconstructed density field with
the group distribution shown in Fig.~\ref{fig_lss}, one can see
clearly the mass concentrations associated with massive
structures, such as the SDSS Great Wall.  Furthermore, one can
also see smaller filaments that are not so evidently seen in the
group distribution itself. For comparison, we also show the
predicted density fields at $z=2$ (middle panel) and $z= 4$ (top
panel). A comparison among the different panels nicely illustrates
the hierarchical formation of the SDSS Great Wall.

Since this paper focusses on the velocity and tidal fields
reconstructed from the SDSS DR7, we refrain from a detailed
quantification of the accuracy of this reconstruction method. A
detailed test of this method against $N$-body simulations, as well
as an in-depth comparison with the reconstruction method proposed
in W09, will be presented in a forthcoming paper. For the moment
we only emphasize that the new method presented here, albeit
simpler and less time-consuming, is unable, by construction, to
resolve highly non-linear structures. Hence, unlike the W09
method, it is only able to reconstruct density field smoothed on
relatively large scales. Its advantage, however, is that it
automatically yields the reconstructed density field {\it as a
function of time}, therefore providing insight into the
merger/assembly history of the large-scale structure that hosts
the galaxies in the SDSS DR7.

\section{Discussion and Summary}

W09 have developed a method to reconstruct the velocity and tidal
fields from the distribution of dark matter haloes. In this paper, we
use the method to reconstruct these two fields in the Sloan Digital
Sky Survey (SDSS) DR7 volume from dark matter halos represented by
galaxy groups. We use detailed mock catalogues to test the reliability
of our method against uncertainties arising from the inaccuracies of
our method, redshift distortions, survey boundaries and false
identifications of groups by our group finder. We find that both the
velocity and tidal fields can be reliably reconstructed in the inner
region of the survey volume, but that the reconstruction near the
survey boundaries is significantly biased, especially the velocity
field, which is more sensitive to large scale structure. We define a
quantity to quantify the closeness to the survey boundary, and find
that for the SDSS DR7 the bias produced by the boundary effects
becomes comparable to or smaller than that produced by other factors
for the inner $\sim 66\%$ of the survey volume. The total bias in the
reconstruction quantities is small, and detailed analysis suggest that
each of the factors mentioned above contributes roughly equally to the
bias. These results show that our method can be applied to real data
to get reliable results.

We apply our method to the galaxy group catalogue obtained from the
SDSS DR7 using the halo-based group finder of Yang \etal (2005). We
use the reconstructed tidal field to classify the morphologies of
large scale structures, based on the number of positive eigenvalues of
the tidal tensor. This reveals clearly a cosmic web, with filamentary
structures connecting clusters and enveloped by sheet-like structures
that surround large voids. In particular, the volume filling fractions
of the four structures are in good agreement with the simulations. As
examples, we show the velocity fields in the region of a large void, a
large region covering the SDSS Great Wall, and in the neighborhood of
a small filament. In agreement with expectations, the velocity fields
are clearly divergent in the centers of voids, while converging
towards sheets (along one direction), filaments (along two directions)
and clusters (along all three directions). The distribution of the
resulting velocities has an approximate Gaussian core with extended
wings, consistent with model expectations. The Gaussian cores for both
the $X$ and $Y$ components peak roughly at zero, with a dispersion of
about 360$\kms$, in good agreement with prediction for the
$\Lambda$CDM concordance cosmology. However, the distribution of the
$Z$ component peaks at about $-117\kms$ and has a larger dispersion of
$413\kms$. This suggests that a large fraction of the entire SDSS
survey volume (equivalent to a sphere with a radius of $\sim 170\mpc$)
is undergoing a bulk flow of $\sim 120\kms$.  Based on the direction
of this bulk flow, it is most likely due to the gravitational
attraction from massive structures, including the SDSS Great Wall,
located at low declination.

We have also used the reconstructed velocity field, together with the
Zel'dovich approximation, to reconstruct the cosmic density field in
the SDSS survey volume. Visual inspections show that this method can
well reproduce both the massive structures populated by rich groups,
and small structures populated by individual galaxies. However, since
it is based on the Zel'dovich approximation, it is unable to resolve
highly non-linear regions, unlike the method presented in W09, which
is based on an analog of the `halo model' (e.g., Cooray \& Sheth 2002)
that describes the matter density distribution (in a statistical
sense) in terms of its halo building blocks. Nevertheless this
reconstruction has the advantage of being able to trace the density
field across cosmic times, thus providing insight into the assembly of
the large scale structure that hosts the SDSS galaxies.

The reconstructed velocity, tidal and density fields presented here
have many applications. For instance, together with the SDSS DR7 group
catalogue, our reconstructed cosmic fields can be used to investigate
how galaxy properties correlate with their environment. In particular,
it would be interesting to investigate whether galaxies in a halo of
given mass $M$ that is located in a filament differ, in a statistical
sense, from galaxies in haloes of the same mass but located in a void,
cluster or sheet. Recent studies have suggested that the properties of
dark matter halos do depend significantly on their large-scale
environments (e.g.  Lee \& Pen 2001; Gao et al. 2005; Wechsler et
al. 2006; Jing et al. 2007; Wang et al. 2007, 2011), and it remains to
be seen whether this is also the case for galaxies.  The reconstructed
density, velocity and tidal fields presented here, together with the
intrinsic properties observed for SDSS galaxies, will provide a unique
and novel avenue to study how environmental effects affect the
properties of galaxies.

The fact that our reconstructed velocity field, together with the
Zel'dovich approximation, seems able to yield a reliable density
field on large, quasi-linear scales also indicates that it can be
used to set up the initial conditions for $N$-body simulations of
the large-scale structure formation in the local universe. In
practice, the velocity field has to be smoothed on a suitable
scale such that strongly nonlinear structures are absent and the
Zel'dovich approximation is valid. The missing part of the
perturbation spectrum in the initial conditions on small scales
can be included using the method developed by, e.g., Bertschinger
(1987), Hoffman \& Ribak (1992), and van de Weygaert \&
Bertschinger (1996).  These initial conditions can then be used to
run constrained simulations that closely mimic the true large
scale structure in the SDSS survey volume (see also e.g. Nusser \&
Dekel 1992; Kolatt et al.  1996; Klypin et al. 2003; Kitaura \&
Ensslin 2008; Forero-Romero et al. 2011). Our reconstructed
density field based on the Zel'dovich approximation demonstrates
the potential of this method, but further investigation along this
line is needed. The hope is that eventually the formation history
of the local universe can be traced back in time with reasonable
accuracy. A comparison with the observed galaxy population within
the same volume will then provide a goldmine to explore how
galaxies form and evolve.

The reconstructed density and velocity fields will also be useful
for studying the dynamics and physics of the IGM. For example, the
line-of-sight peculiar velocities of the most massive groups in
the SDSS DR7 survey volume can be used to make detailed
predictions for the kinetic Sunyaev-Zel'dovich effect, which can
be compared to forthcoming observations from, e.g., the Atacama
Cosmology Telescope (ACT; Swetz et al. 2011) and the Planck
satellite. In addition, a comparison of the reconstructed density
field with quasar absorption line studies that are sensitive to
the absorptions in the SDSS DR7 survey volume can provide
invaluable constraints on the temperature and metallicity of the
filaments and sheets that make up the cosmic web. Observations
have so far revealed a wide array of absorption lines, ranging
from low ions such as HI all the way up to highly ionized species,
such as OVI and NeVIII (Tripp \& Bowen 2006 and references
therein), presumably associated with the warm-hot medium seen in
gas-dynamical simulations (e.g. Cen \& Ostriker 1999; Dav\'e et
al. 2001).  With the installation of the {\it Cosmic Origins
  Spectrograph} ({\it COS}) on the {\it HST}, the sample of UV
absorption systems in the local universe should increase by an
order-of-magnitude or more.  Such observations, together with the
information about the density and velocity fields obtained from our
reconstruction, provide an unique avenue to understand the nature of
absorption systems at low-$z$ and their implications for the state and
structure of the IGM at the present time, and in particular to explore
the connection and interaction between the IGM and the galaxy
population.

Our reconstructed velocity field can also be used to constrain
cosmological parameters via a comparison with the peculiar velocity
field obtained from distance indicators (e.g. Kaiser et al. 1991;
Willick \& Strauss 1998; Hudson et al. 2004; Colombi et al. 2007). It
should be kept in mind that our reconstruction method is cosmology
dependent: cosmology enters through the method we use to assign halo
masses to our galaxy groups (see Y07 for details), and via the bias
parameter, $b_{\rm hm}$ (Eq.~[\ref{eq:bhm}]) and the cosmological
parameters $\Omega_{\rm m}$ and $H_0$ that enter the reconstruction of
the velocity field. Since methods that rely on distance indicators are
less cosmology-dependent, a comparison of the velocity fields obtained
using both methods may be able to constrain one or more cosmological
parameters. Previous studies usually used the galaxy distribution
directly to compare to the velocity field.  However, galaxies are
known to be biased tracers of the large scale mass distribution and
the exact form of this bias is complicated, as it depends on various
properties of the galaxies, such as luminosity and color. Thus using
galaxies directly may introduce uncertainties into the estimate of
cosmological parameters. Our reconstructed velocity field is a better
choice for this purpose, because the bias of dark matter halos is well
understood. In particular, it is interesting to investigate if the
large-scale bulk flow revealed in our analysis might pose a challenge
to the standard $\Lambda$CDM model.

The reconstructed density, velocity and tidal fields presented here
are publicly available from the authors upon request. We hope that
they will provide a useful piece of data for investigations in
cosmology, large-scale structure, galaxy formation, and the IGM.

\section*{Acknowledgments}

This work is supported by NSFC (11073017, 10821302, 10925314,
11128306), 973 program (2007CB815402, 2009CB824800) and the
Fundamental Research Funds for the Central Universities. HJM would
like to acknowledge the support of NSF AST-0908334 and the
CAS/SAFEA International Partnership  Program for Creative Research
Teams (KJCX2-YW-T23).

\newpage

\end{document}